\documentclass[iop]{emulateapj}

\slugcomment{Accepted for publication in the  Astrophysical Journal}

\shorttitle{Chemical abundances of the Bo\"otes~II  dwarf spheroidal}
\shortauthors{A. Koch \& R.M. Rich}

\begin{document}

\title{A chemical confirmation of the faint Bo\"otes~II dwarf Spheroidal Galaxy}

\author{
Andreas Koch\altaffilmark{1},   
\& R. Michael Rich\altaffilmark{2}
}
\altaffiltext{1}{Landessternwarte, Zentrum f\"ur Astronomie der Universit\"at Heidelberg, K\"onigstuhl 12, 69117 Heidelberg, Germany}
\altaffiltext{2}{University of California Los Angeles, Department of Physics \& Astronomy, Los Angeles, CA, USA}
\email{akoch@lsw.uni-heidelberg.de}
\begin{abstract}
We present a chemical abundance study of the brightest confirmed member star of the ultrafaint dwarf galaxy Bo\"otes II from Keck/HIRES high-resolution 
spectroscopy at moderate signal-to-noise ratios. 
{At  [Fe/H] = $-2.93\pm0.03$(stat.)$\pm0.17$(sys.) this star 
chemically resembles metal-poor halo field stars and the signatures of other faint dwarf spheroidal galaxies at the same metallicities in that it shows 
enhanced [$\alpha$/Fe] ratios, Solar Fe-peak element abundances, and low upper limits on the neutron-capture element Ba.
Moreover, this star shows no chemical peculiarities in any of the eight elements we were able to measure.}
This implies that the chemical outliers  found in other systems 
remain outliers pertaining to the unusual enrichment histories of the respective environments, while Boo~II
appears to have experienced an enrichment history typical of its very low mass.
We also re-calibrated previous measurements of the galaxy's metallicity from the calcium triplet (CaT) and find a much lower value 
than reported before. The resulting broad metallicity spread, in excess of one  dex, the very metal poor mean, and the 
chemical abundance patterns of the present star  imply that Bo\"otes~II is a low-mass, old, metal poor dwarf galaxy and 
not an overdensity associated with the Sagittarius Stream as has been previously suggested based on its sky position and kinematics. 
The low, mean CaT metallicity  of $-$2.7 dex falls right on the luminosity-metallicity relation delineated over four orders of magnitude from the  more luminous to the faintest 
galaxies. Thus Bo\"otes~II's chemical enrichment appears representative of the galaxy's original mass, while tidal stripping and other mass loss mechanisms 
were probably not significant as for other low-mass satellites.
\end{abstract}
%
%
%
%
\section{Introduction}
Within the last three years, the galaxy luminosity function has been vastly extended to include a new population of ultra-faint (UF) satellites, often with total magnitudes 
M$_V< -6$ (e.g., Willman et al. 2005; Belokurov et al. 2007). At 10$^3$--10$^5$ M$_{\odot}$ their stellar masses are comparable to the most extended Milky Way star clusters (Gilmore et al. 2007). 
In fact, for some of the UF systems it is even unclear whether they
  are truly old, metal-poor systems like the dwarf spheroidal (dSph) galaxies, or whether they are extended, disrupting, essentially dark-matter free star clusters (e.g., 
 Niederste-Ostholt et al. 2009; cf. Simon et al. 2011; Belokurov et al. 2014).
In contrast, for the cases where the measured velocity dispersions indicate mass-to-light  ratios of up to several hundred, these galaxies appear to be the most dark matter dominated stellar systems known in the
  universe (e.g., Gilmore et al. 2007; Simon et al. 2011).

Early high dispersion spectroscopy of the more luminous dSph galaxies found low [$\alpha$/Fe] ratios   relative to field stars in the halo (e.g., Shetrone et al. 2001; Venn et al. 2004; 
Tolstoy et al. 2009), which was first interpreted as strict evidence against a simplistic building-block scenario, according to which the Galactic halo was formed through the 
accretion of dSph-like objects at early times (Searle \& Zinn 1978). 
However, recent high dispersion analyses found that there is 
in fact an overlap of the dSph abundances with the halo abundance patterns at the metal-poor end (e.g., Cohen  \& Huang 2010; Frebel et al. 2010; Gilmore et al. 2013). Recent studies 
also detected an emerging population with [Fe/H]$<-3$ in the UF and the luminous dSphs (e.g., Tafelmeyer et al. 2010), 
{which traditionally have been claimed to be missing in the classical dwarf galaxies (Helmi et al. 2006)}.
Altogether, this evidence makes the UF dSphs promising candidates to have contributed to the build-up of the metal-poor Galactic halo.  

Likewise, halo stars with peculiar chemical element patterns 
{may have originated from} the dSphs, as the latter also show a number of intriguing peculiarities (e.g., Koch et al. 2008; Simon et al. 2010). Examples include extraordinarily high 
[Mg/Ca] ratios, which suggest stochastic enrichment by very massive stars and thus an unprecedented mode of star formation acting in those low-mass galaxies. 
The [Co/Cr] ratios in some UF dSph stars are also very high, consistent with these galaxies hosting the very first stars to form (Population~III).  
Some dSph stars with [Fe/H]$<-2$ show little or no measurable 
neutron-capture elements  (Fulbright et al. 2004; Koch et al. 2008; Roederer \& Kirby 2014). 
 Therefore, the element distributions in the UF dSphs hold important keys to the understanding of 
both stellar and galactic evolution: what governed the stochastic  sampling of the initial mass function in such systems? What do those patterns reveal about the yields of the contributing 
supernovae (SNe) and thus the nucleosynthetic origin of the elements?

Regrettably,  abundance ``patterns'' are only known for seven of the $\sim$15 currently known UF dSphs  (McConnachie 2012), 
often with only a few stars observed per galaxy and, in some cases, only for a limited range of elements.  
Nonetheless, every 
{dSph galaxy}
 studied so far has revealed peculiar chemical abundance ratios of some sort, 
even if only one star  (usually the brightest giant) was reachable with time-intensive high-resolution spectroscopy.

Here, we present high dispersion 
spectra of one star  in Bo\"otes~II (herafter Boo~II), which at M$_V=-2.7$ is one of the least luminous galaxies  studied to date  (Walsh et al. 2007, 2008).  Since these are low
 luminosity systems, their stellar populations are sparse and giants on the upper red giant branch (RGB) are rare.  Consequently, we are running out of stars bright enough to observe at 
 high resolution, even with 8--10 m telescopes and we must turn to the lower RGB, where only five confirmed members are known in Boo II. 

Boo~II has been confirmed kinematically as a stellar system by Koch et al. (2009a; hereafter K09).
Yet, its nature is far from understood, since its distance, radial velocity, and 
position on the sky coincide with that of the Sagittarius (Sgr) stream (Fellhauer et al. 2006). However, the low-resolution, low signal-to-noise (S/N) metallicity measurements of K09 
show a metal-poor mean and hint at an abundance spread  
that would favor Boo~II being a dark matter dominated 
system, not a disrupted Sgr cluster, which was bolstered by its velocity dispersion of $\sim 10$ km\,s$^{-1}$.
Moreover,  Boo~II was found to be  more metal-rich (K09) than its low luminosity would imply 
if it was to follow the mass-metallicity relation that appears to hold over several orders of magnitude (e.g., Kirby et al. 2011). 
Clearly, a full-fledged investigation of Boo~II's abundance pattern is needed to better understand its nature and its evolutionary and chemical enrichment characteristics. 
\section{Spectroscopic data}
The target selected for this study was star 15 (SDSS J135751.18+125136.9), the ``brightest'' object in the member list of K09. 
At V=19 mag it lies well below the tip of the RGB, indicated by the isochrones in Fig.~1,  
and below the horizontal branch (HB) level. However, no photometric study of Boo~II has so far revealed  
any brighter RGB population in this faint stellar system (e.g., Walsh et al. 2007, 2008). 
\begin{figure}[htb]
\begin{center}
\includegraphics[angle=0,width=1\hsize]{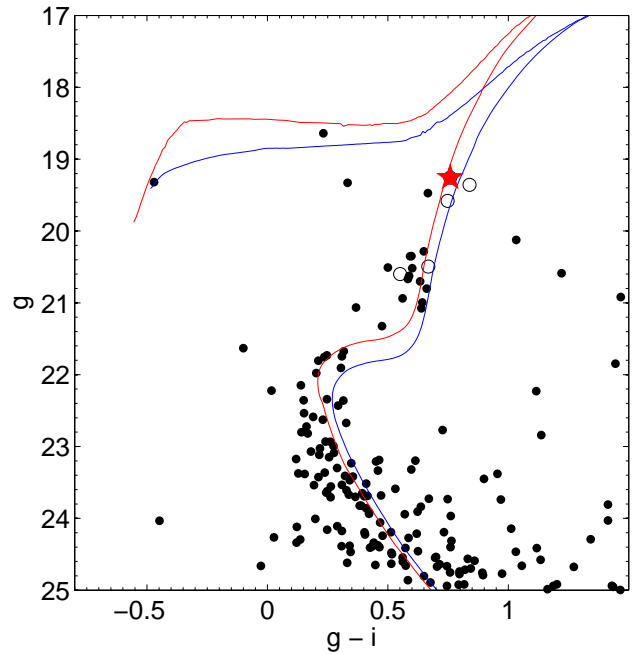}
\end{center}
\caption{CMD of stars within the central 3' of Boo~II, using the photometry of K09. The target of this study is highlighted as a red star symbol. 
Open circles denote the radial velocity members of K09. Also shown are two metal-poor ($-1.8$ and $-2.5$ dex), old (12 Gyr), $\alpha$-enhanced Dartmouth isochrones (Dotter et al. 2008), shifted to the distance 
modulus of Boo~II as derived by K09.}
\end{figure}
This star is located 3.4' from the nominal center of the galaxy, which compares to its half-light radius of $\sim$4.0' (Martin et al. 2008; K09)\footnote{Note, however,  
that Walsh et al. (2007) reported a smaller radius of $r_h$=2.5'.}. 
Some characteristic data for Boo~II-15 are given in Table~1. 
\begin{center}
\begin{deluxetable}{rc}[htb]
\tabletypesize{\scriptsize}
\tablecaption{Properties of Boo~II-15}
\tablewidth{0pt}
\tablehead{
\colhead{Parameter} & \colhead{Value}
}
\startdata
 $\alpha$ (J2000.0)	     & 13:57:51.2 \\
 $\delta$ (J2000.0)	     & 12:51:36.6 \\
 $g$\tablenotemark{a}  [mag] & 19.26 \\
 $r$\tablenotemark{a}  [mag] & 18.76 \\
 $i$\tablenotemark{a}  [mag] & 18.50 \\
 V\tablenotemark{b}  [mag]   & 18.98 \\
 B$-$V\tablenotemark{b} [mag]& 0.65 \\
 T$_{\rm eff}$(B$-$V)  [K]   & 5264 \\
 T$_{\rm eff}$(spec.)  [K]   & 5000 \\
 log\,$g$ 		     & 2.26 \\
 $\xi$ [km\,s$^{-1}$]	     & 1.81
\enddata
\tablenotetext{a}{INT/WFC photometry of K09.}
\tablenotetext{b}{Obtained from the 
 $g$,$r$,$i$
magnitudes 
using the transformations of Jordi et al. (2006).}
\end{deluxetable}
\end{center}

Our data were taken on May 23, 2014, using the High Resolution Echelle Spectrometer (HIRES; Vogt et al. 1994) on the Keck telscope.
The instrument was operated with the C5 decker (width 1.15''), which yields a spectral resolving power of $R\sim36000$ and a full wavelength range
of 4300--8350~\AA. 
Conditions were photometric with fair seeing of  0.7''--1.1''; we obtained a total
of  8$\times$2700 s or 6 hours of  exposure time.

The raw data were reduced within the MAKEE software package\footnote{MAKEE was developed by T. A. Barlow specifically for reduction of Keck HIRES data. 
It is freely available on the World Wide Web at the Keck Observatory home page, \tt{http://www2.keck.hawaii.edu/inst/hires/makeewww}} 
data reduction package, which performs standard steps such as bias- and flat field correction, wavelength calibration via Th-Ar arc lamps, 
optimal extraction, and sky subtraction. 
The resulting spectra have a signal-to-noise (S/N) ratio of up to 40 per pixel near H$\alpha$, degrading to $\sim$20--25 near 4500~\AA. 
Finally, the membership of star 15 with Boo~II was independently confirmed by remeasuring its radial velocity via cross-correlation  against a synthetic RGB spectrum using IRAF's {\em fxcor}.
{Thus, our current value of $-104.8$ km\,s$^{-1}$ compares to the heliocentric radial velocity of $-100.1\pm2.3$ km\,s$^{-1}$ measured in K09. Potential small-scale variations in the 
velocities are beyond the scope of the present analysis (cf. Koch et al. 2014).}
\section{Abundance analysis}
In full analogy to previous works (e.g., Koch et al. 2009b), we performed a standard equivalent width (EW) analysis to derive 
the chemical element abundance ratios of Boo~II-15, with the exception for C,  
for which we used spectral synthesis. 

To this end, we used the Kurucz atmosphere
grid\footnote{{\tt http://kurucz.harvard.edu}}, operating in local
thermodynamic equilibrium (LTE), without convective overshoot, and
using the $\alpha$-enhanced opacity distributions AODFNEW (Castelli \&
Kurucz 2003)\footnote{{\tt http://wwwuser.oat.ts.astro.it/castelli}}.
While we have no {\em a priori} knowledge of the actual level of $\alpha$-enhancement in this star, 
measurements of other metal-poor, faint dSphs indicate a broad scatter in the [$\alpha$/Fe] ratios and 
a partial overlap with Galactic halo stars, at [$\alpha$/Fe$]\sim$0.4 dex. 
However, our investigation of   
systematic uncertainties (Sect.~3.3) showed that the choice of enhanced (AODFNEW) or Solar (ODFNEW) opacity distributions 
has no impact on the derived abundance ratios in excess of 0.02 dex.
\subsection{Line list}
Our line list is the same as used in Koch et al. (2009b, and references therein). EWs were determined by fitting a Gaussian profile to the lines 
using IRAF's {\em splot} task and all our measurements are listed in Table~2. 
The effect of hyperfine splitting was accounted for in our spectral synthesis of the four Ba lines that fall within HIRES' wavelength range (Sect.~3.9)
by using the splitting information of McWilliam et al. (1995). 
\begin{center}
\begin{deluxetable*}{rcccc|rcccc}
\tabletypesize{\scriptsize}
\tablecaption{Linelist}
\tablewidth{0pt}
\tablehead{
\colhead{Ion} & \colhead{$\lambda$ [\AA]} & \colhead{E.P. [eV]} & \colhead{$\log\,gf$} & \colhead{EW [m\AA]\tablenotemark{a}} &
\colhead{Ion} & \colhead{$\lambda$ [\AA]} & \colhead{E.P. [eV]} & \colhead{$\log\,gf$} & \colhead{EW [m\AA]\tablenotemark{a}} 
}
\startdata
C (CH) & \multicolumn{3}{c}{$A^2\Delta - X^2\Pi$ G band} & syn       &    Fe\,{\sc i}  &  5191.465  &  3.030  &    $-$0.656   &    24.6    \\
Na\,{\sc i}   &  5889.973  &  0.000  &  $-$0.462 &  161              &    Fe\,{\sc i}  &  5192.353  &  2.998  &    $-$0.521   &    32.4    \\
Na\,{\sc i}   &  5895.940  &  0.000  &  $-$0.801 &  141              &    Fe\,{\sc i}  &  5194.950  &  1.557  &    $-$2.060   &    41.8    \\
Mg\,{\sc i}   &  4571.096  &  0.000  &  $-$5.691 &  30.9             &    Fe\,{\sc i}  &  5216.280  &  1.608  &    $-$2.120   &    44.7   \\
Mg\,{\sc i}   &  4702.991  &  4.346  &  $-$0.666 &  56.0             &    Fe\,{\sc i}  &  5232.952  &  2.940  &    $-$0.080   &    57.4  \\
Mg\,{\sc i}   &  5172.684  &  2.711  &  $-$0.402 & 165               &    Fe\,{\sc i}  &  5367.480   &  4.420  &  \phs0.440   &  17.0	  \\
Mg\,{\sc i}   &  5183.604  &  2.717  &  $-$0.180 & 191               &    Fe\,{\sc i}  &  5383.370  &  4.310  &  \phs0.650   &   28.4  \\
Mg\,{\sc i}   &  5528.418  &  4.346  &  $-$0.481 &  55.6             &    Fe\,{\sc i}  &  5397.141  &  0.914  &    $-$1.993   &   91.4	\\
Ca\,{\sc i}   &  4318.650  &  1.900  &  $-$0.210 &  50.3             &   Fe\,{\sc i}  &  5405.785  &  0.990  &    $-$1.844   &   92.4	\\
Ca\,{\sc i}   &  5588.764  &  2.520  &\phs0.310 &  29.8              &   Fe\,{\sc i}  &  5424.080  &  4.320  &  \phs0.580   &	16.8   \\
Ca\,{\sc i}   &  6102.727  &  1.880  &  $-$0.790 &  22.3             &   Fe\,{\sc i}  &  5429.706  &  0.914  &    $-$1.879   &   98.6  \\
Ca\,{\sc i}   &  6122.226  &  1.890  &  $-$0.320 &  50.0             &   Fe\,{\sc i}  &  5434.534  &  1.011  &    $-$2.122   &    81.5 \\
Ca\,{\sc i}   &  6162.180  &  1.899  &  $-$0.090 &  46.6             &   Fe\,{\sc i}  &  5446.924  &  0.990  &    $-$1.930   &    98.0   \\
Ca\,{\sc i}   &  6439.083  &  2.526  &\phs0.390 &  35.0              &   Fe\,{\sc i}  &  5497.526  &  1.011  &    $-$2.840   &    40.1 \\
Ca\,{\sc i}   &  6493.780  &  2.520  &\phs0.020 &  19.5              &   Fe\,{\sc i}  &  5501.477  &  0.950  &    $-$2.950   &    43.0	\\
Ti\,{\sc i}   &  4991.070  &  0.836  &\phs0.380 &  22.0              &   Fe\,{\sc i}  &  5506.791  &  0.990  &    $-$2.797   &    45.9  \\
Ti\,{\sc i}   &  4999.510  &  0.826  &\phs0.140 &  26.0              &   Fe\,{\sc i}  &  5569.631  &  3.417  &    $-$0.490   &    14.3  \\
Ti\,{\sc i}   &  5210.390  &  0.048  &  $-$0.884 &  13.4             &   Fe\,{\sc i}  &  5572.851  &  3.396  &    $-$0.280   &    20.9  \\
Ti\,{\sc ii}  &  4417.719  &  1.165  &  $-$1.430 &  52.4             &   Fe\,{\sc i}  &  5586.771  &  3.368  &    $-$0.120   &    35.0	\\
Ti\,{\sc ii}  &  4443.794  &  1.080  &  $-$0.700 &  74.1             &   Fe\,{\sc i}  &  5615.658  &  3.332  &  \phs0.050   &	  45.0	\\
Ti\,{\sc ii}  &  4468.507  &  1.131  &  $-$0.600 &  69.5             &   Fe\,{\sc i}  &  6136.624  &  2.453  &    $-$1.400   &    48.0	\\
Ti\,{\sc ii}  &  4501.273  &  1.116  &  $-$0.750 &  69.5             &   Fe\,{\sc i}  &  6137.000  &  2.198  &    $-$2.950   &    10.0	\\
Ti\,{\sc ii}  &  4563.761  &  1.221  &  $-$0.960 &  66.3             &   Fe\,{\sc i}  &  6137.702  &  2.588  &    $-$1.403   &    34.0   \\
Ti\,{\sc ii}  &  4571.968  &  1.572  &  $-$0.530 &  66.5             &   Fe\,{\sc i}  &  6191.571  &  2.432  &    $-$1.420   &    44.0	  \\
Ti\,{\sc ii}  &  5226.545  &  1.566  &  $-$1.300 &  34.0             &   Fe\,{\sc i}  &  6230.730  &  2.560  &    $-$1.281   &    43.0	  \\
Cr\,{\sc i}   &  5206.044  &  0.941  &\phs0.019 &  50.0              &   Fe\,{\sc i}  &  6252.565  &  2.404  &    $-$1.720   &    20.4   \\
Cr\,{\sc i}   &  5409.799  &  1.004  &  $-$0.720 &  15.0             &   Fe\,{\sc i}  &  6393.612  &  2.430  &    $-$1.430   &    26.5    \\
Fe\,{\sc i}  &  4903.316  &  2.882  &   $-$1.080   &	24.0         &   Fe\,{\sc i}  &  6400.000  &  3.603  &   $-$0.520   &	   17.8    \\
Fe\,{\sc i}  &  4918.998  &  2.850  &   $-$0.672   &    52.7	     &   Fe\,{\sc i}  &  6494.994  &  2.400  &   $-$1.273   &	   34.0	   \\
Fe\,{\sc i}  &  4994.138  &  0.914  &   $-$3.080   &    22.0	     &   Fe\,{\sc i}  &  6592.910   &  2.730  &   $-$1.470   &	   14.3   \\
Fe\,{\sc i}  &  5001.870  &  3.881  &  \phs0.010   &    27.0	     &   Fe\,{\sc i}  &  6677.997  &  2.692  &    $-$1.470   &    25.0	 \\
Fe\,{\sc i}  &  5041.760  &  1.485  &   $-$2.086   &    50.6	     &   Fe\,{\sc ii} &  4508.214  &  2.855  &     $-$2.210  &    21.6   \\
Fe\,{\sc i}  &  5049.830  &  2.270  &   $-$1.350   &    35.0	     &   Fe\,{\sc ii} &  4583.837  &  2.806  &     $-$2.020  &    45.0	 \\
Fe\,{\sc i}  &  5051.640  &  0.914  &   $-$2.780   &    43.4	     &   Fe\,{\sc ii} &  4923.930  &  2.891  &     $-$1.320  &    66.1    \\
Fe\,{\sc i}  &  5068.770  &  2.940  &   $-$1.230   &    17.0	     &   Fe\,{\sc ii} &  5197.576  &  3.230  &     $-$2.250  &    11.6    \\
Fe\,{\sc i}  &  5083.340  &  0.958  &   $-$2.910   &    35.6	     &   Fe\,{\sc ii} &  5234.630  &  3.220  &     $-$2.240  &    16.0	 \\
Fe\,{\sc i}  &  5123.730  &  1.011  &   $-$3.068   &    26.4	     &   Ni\,{\sc i}   &  5476.921  &  1.826  &  $-$0.890 &  40.0 \\
Fe\,{\sc i}  &  5127.360  &  0.915  &   $-$3.307   &    26.0	     &   Ba\,{\sc ii}  &  4554.029  &  0.000  &\phs0.170 &  hfs, syn \\
Fe\,{\sc i}  &  5133.680   &  4.180  &  \phs0.200   &    19.0	     &   Ba\,{\sc ii}  &  5853.680  &  0.600  &  $-$1.010 &  hfs, syn \\
Fe\,{\sc i}  &  5150.852  &  0.990  &   $-$3.000   &    30.0	     &   Ba\,{\sc ii}  &  6141.727  &  0.704  &  $-$0.077 &  hfs, syn \\
Fe\,{\sc i}  &  5166.280  &  0.000  &   $-$4.195   &    33.0	     &   Ba\,{\sc ii}  &  6496.908  &  0.600  &  $-$0.380 &  hfs, syn \\
Fe\,{\sc i}  &  5171.610  &  1.480  &    $-$1.760   &    59.2        &                 &            &         & 
\enddata							
\tablenotetext{a}{``HFS'' denotes that hyperfine splitting was accounted for in the analysis of this line, while ``syn'' indicates spectral synthesis.}
\end{deluxetable*}
\end{center}
\subsection{Stellar parameters}
As an initial estimate, we obtained a photometric temperature using the calibration of Alonso et al. (1999), 
which yields T(B$-$V)=5264$\pm$109 K and accounts for  the photometric and the transformation 
uncertainties. 
Here, we adopted a reddening of E(B$-$V)=0.03 from the Schlafly \& Finkbeiner (2011) maps and
used the transformation of our $g$ and $i$ photometry to the Johnson-Cousins system provided by Jordi et al. (2006). 

As a refinement, we obtained a spectroscopic temperature by enforcing excitation equilibrium, i.e., 
by removing the trend of abundance from neutral iron lines with excitation potential.  
Similarly, the microturbulence was fixed by removing the slope in the plot of  the reduced width, RW=$\log({\rm EW}/\lambda$).
Weak lines with $\log({\rm EW}/\lambda)<-5.5$ were removed from this exercise. 
The precision on temperature and microturbulence, under which 
reasonably flat slopes could still be achieved, is on the order of 
150 K and 0.15 km\,s$^{-1}$. 
This also renders the discrepancy between the photometric and the spectroscopic temperatures of 250 K insignificant.

The surface gravity of Boo~II-15 was set by minimizing the ionization imbalance between neutral and ionized iron lines. 
Owing to the blue extent of our spectrum, we could ensure the  inclusion of a sufficient number of Fe\,{\sc ii} lines.
Thus ionization equilibrium could be satisfied to within  [Fe\,{\sc i}/Fe\,{\sc ii}]$=0.04\pm0.06$ dex.
This was also confirmed by a good balance between the abundances of neutral and ionized Ti, in that $\log\,\varepsilon$(Ti\,{\sc i})$-\log\,\varepsilon$(Ti\,{\sc ii})$=-0.10\pm0.13$ dex.
All parameters were iterated towards convergence {and we list the final adopted values in Table~1}. 
\subsubsection{Photometric parameters}
{Since the derivation of stellar parameters from excitation and ionization balance may be susceptible to NLTE effects  in  metal-poor stars
we also investigated the effect of using only photometric parameters in the following (e.g., Norris et al. 2010a,b; Roederer \& Kirby 2014). 
If we employ the basic stellar equations (e.g., eq.~4 in Bensby et al. 2003), the spectroscopic T$_{\rm eff}$ yields the same 
log\,$g$ that we also obtained from ionization equilibrium, viz. 2.26 dex. 
Upon using T(B$-$V) as an input, we find a gravity that is 0.12 dex higher, at log\,$g$=2.39.
In this approach  we adopted a distance to Boo~II of 46 kpc (K09) and the typical distance error of
4 kpc translates into an uncertainty on log\,$g$ of 0.08 dex. 

Secondly, we consulted the same isochrones as shown in Fig.~1 (Dotter et al. 2008)
to evaluate the best-fit photometric parameters of Boo~II-15. 
This way we obtain a temperature and gravity of 5160 K and 2.33 dex, which are similar to the spectroscopically determined values.
The above tests also justify the order of magnitude of uncertainties we adopted in the following error analysis. 

The effects that using these photometric parameters would have on the abundance ratios in terms of 
changes in T$_{\rm eff}$ and log\,$g$ can be estimated from  Table~3 and the discussions in Sect.~3.3.
Essentially, switching to the isochrone-based parameters renders this star more metal rich (in Fe\,{\sc i}) by 0.17 dex.  
In the following we proceed by using the spectroscopic  T$_{\rm eff}$ and log\,$g$. 
}
\subsection{Abundance errors}
All resulting abundance ratios are listed in Table~3.
These results employed the Solar abundances of Asplund et al. (2009). 
Here, we also indicated the standard deviation and number of lines that were used. 
\begin{center}
\begin{deluxetable*}{crcccccccc}
\tabletypesize{\scriptsize}
\tablecaption{Chemical abundance ratios and systematic errors of Boo~II-15}
\tablewidth{0pt}
\tablehead{\colhead{Ion} & \colhead{[X/Fe]}  & \colhead{$\sigma$} & \colhead{N} 
 & \colhead{T$_{\rm eff}\pm$150 K}  & \colhead{log\,$g\pm$0.2 dex}  &  \colhead{[M/H]$\pm$0.2 dex}  & \colhead{$\xi\pm$0.15 km\,s$^{-1}$} & \colhead{ODF} & \colhead{$\sigma_{\rm tot}$}
}
\startdata
$[$Fe\,{\sc i}/H]              & $-$2.93 &    0.17 &    46 &  $\pm$0.17 &   $<$0.01 &   $<$0.01 & $\pm$0.02 & \phs0.01 & 0.17 \\
$[$Fe\,{\sc ii}/H]             & $-$2.97 &    0.12 &     5 &  $\pm$0.01 & $\mp$0.06 &   $<$0.01 & $\pm$0.02 &  $<$0.01 & 0.09 \\
$[$C\,{\sc i}/Fe\,{\sc i}]     &    0.03 & \nodata & synth & $\pm$0.30 & $\mp0.08$ & $\mp$0.18 &  $<$0.01 & $-$0.35 & 0.37  \\
$[$Na\,{\sc i}/Fe\,{\sc i}]\rlap{$_{\rm NLTE}$}    &    0.15 &    0.02 &     2 & $\pm$0.16 & $\mp$0.01 & $\pm$0.01 & $\pm$0.10 & \phs0.01  & 0.19  \\
$[$Mg\,{\sc i}/Fe\,{\sc i}]     		   &	0.58 &    0.11 &     5 & $\pm$0.16 & $\mp$0.03 &   $<$0.01 & $\pm$0.03 & \phs0.01  & 0.17  \\
$[$Ca\,{\sc i}/Fe\,{\sc i}]     		   &	0.37 &    0.14 &     7 & $\pm$0.11 & $\mp$0.01 &   $<$0.01 & $\pm$0.01 & \phs0.01  & 0.12  \\
$[$Ti\,{\sc i}/Fe\,{\sc i}]     		   &	0.27 &    0.17 &     3 & $\pm$0.19 & $\mp$0.01 &   $<$0.01 &   $<$0.01 & \phs0.02  & 0.21  \\
$[$Ti\,{\sc ii}//Fe\,{\sc ii}]  		   &	0.23 &    0.21 &     7 & $\pm$0.06 & $\mp$0.06 & $\pm$0.01 & $\pm$0.05 &  $<$0.01  & 0.12  \\
$[$Cr\,{\sc i}/Fe\,{\sc i}]     		   & $-$0.38 &    0.03 &     2 & $\pm$0.18 & $\mp$0.01 &   $<$0.01 & $\pm$0.02 & \phs0.01  & 0.18  \\
$[$Ni\,{\sc i}/Fe\,{\sc i}]     		   & $-$0.05 & \nodata &     1 & $\pm$0.17 &   $<$0.01 &   $<$0.01 & $\pm$0.02 & \phs0.01  & 0.18  \\
$[$Ba\,{\sc ii}/Fe\,{\sc ii}]   		   & $<$0.62 & \nodata & 4 & $\pm$0.11 & $\mp$0.06 & $\pm$0.01 & $\pm$0.01 & $<$0.01 & 0.13 
\enddata
\end{deluxetable*}
\end{center}

In order to assess the systematic uncertainties, we proceeded in the standard manner of 
varying each stellar parameter individually by its errors (Sect.~3.2) and re-deriving all abundance ratios.
The resulting deviations from the abundances obtained from the fiducial parameter set are given in Table~3. 
Here, the influence of the $\alpha$-enhancement of the atmospheres was tested by using both the enhanced and Solar 
 opacity distribution functions, AODFNEW vs. ODFNEW and the resulting difference is given in the column ``ODF''.
Since the actual uncertainty on the [$\alpha$/Fe] ratio
 will be smaller than the difference between both ODFs  of 0.4 dex, the contribution of this source to the final error will be
typically lower by a factor of $\sim$4. 

{For those cases, where only one transition was measurable, no line-to-line scatter can be stated. 
For most of the other elements this statistical error 
amounts to 0.05 dex, which can be thought of as representative for the species with only one line as well. 
It is important to note, then, that this component is in any case small compared to the systematic error}.
As a measure for the overall error on our abundance ratios, we provide $\sigma_{\rm tot}$ as the individual contributions, added in quadrature, 
which neglects possible covariances between the parameters and should thus be considered as a conservative upper limit. 
\subsection{Iron abundance}
We find an iron abundance for Boo~II-15 of [Fe/H]$=-2.93\pm0.03$ dex, rendering it a typical 
representative of the recently detected and analysed metal-poor, very faint dSph galaxy population (e.g., Norris et al. 2010a; Frebel et al. 2010; Ad\'en et al. 2011; Roederer \& Kirby 2014).
At these low metallicities, there is also a vast overlap with the very-to-extremely metal-poor halo stars (e.g., Beers \& Christlieb 2005). 
The high-resolution measurement of [Fe/H] from the present work is considerably lower than our previous 
metallicity estimate from the near-infrared calcium triplet (CaT) lines, reported in  K09 as $-1.81\pm0.10$.
The latter spectra were obtained with a S/N ratio similar to our HIRES data, but at a spectral resolving power that was 
ten times lower than for the present high-resolution spectra. 
 
Several factors complicate the determination of accurate metallicities from the CaT at the low metallicities that the present analysis implies. 
Firstly, all targets of K09 were on the lower RGB below the HB level. 
The widely used calibrations of the line strength of the CaT onto metallicity is generally parameterized in terms of the stellar magnitude 
{\em above} the HB level (V$-$V$_{\rm HB}$), which effectively acts as a gravity indicator (e.g., Armandroff \& Zinn 1988; Rutledge et al. 1997, and references therein).
As several recent works have shown, the linear dependence of line strength on this photometric gravity indicator
levels off further below the HB (e.g., Da Costa et al. 2009). 

Secondly, the classical calibrations of the CaT onto metallicity relied on linear expressions
that had been established by means of globular cluster (GC) reference scales of known metallicities. 
As a consequence, every calibration is strictly only valid over the metallicity range covered by the calibration GCs. 
We will revisit the question of these calibrations in Sect.~4.
\subsection{Carbon abundance}
The C-abundance of Boo~II was found by spectral synthesis of the CH G-band near 4300~\AA, which is included at the blue end of our spectra. 
This synthesis used a  line list kindly provided by B. Plez (priv. comm.). 
The resulting Solar [C/Fe] ratio is fully compatible with C-normal very metal-poor stars in the Galactic halo and UF dSphs at comparably low metallicity (e.g., Aoki et al. 2007; Norris et al. 2010a; Yong et al. 2013) 
without the need to invoke C-production and/or transfer mechanisms (e.g., in binaries) as for the population of Carbon-enhanced Extremely metal-poor (CEMP)
stars that is populous in the halo and some dSphs (e.g., Beers \& Christlieb 2005; Aoki et al. 2007; Norris et al. 2010a,c; Masseron et al. 2010; Lee et al. 2013). 
We note that the adopted O-abundance, which is of importance for the molecular equilibrium employed in our atmospheres, has only little impact on 
the resulting [C/Fe] ratio in that a one dex change in [O/Fe] incurs a mere 0.05 dex variation in [C/Fe]. 

Fig.~2 shows the [C/Fe] abundance ratio, and similarly for other chemical elements, for Boo~II-15 in the context of other small-scale stellar systems.
We compare our data to the recent sample of 313 metal-poor halo stars that were homogeneously analysed by Roederer et al. (2014). 
Also shown are the, still limited, data for high-resolution abundance ratios in those seven of the UF dSphs studied to date, comprising
$\sim$40 stars in   Hercules (Koch et al. 2008, 2013; Ad\'en et al. 2011); Boo~I (Feltzing et al. 2009; Norris et al. 2010b; Gilmore et al. 2013; Ishigaki et al. 2014); 
Leo~IV (Simon et al. 2010); Coma Ber and Ursa Major II (Frebel et al. 2010); Segue~1 (Norris et al. 2010c; Frebel et al. 2014); and Segue~2 (Roederer \& Kirby 2014).
\begin{figure*}[!ht]
\begin{center}
\includegraphics[angle=0,width=0.4885\hsize]{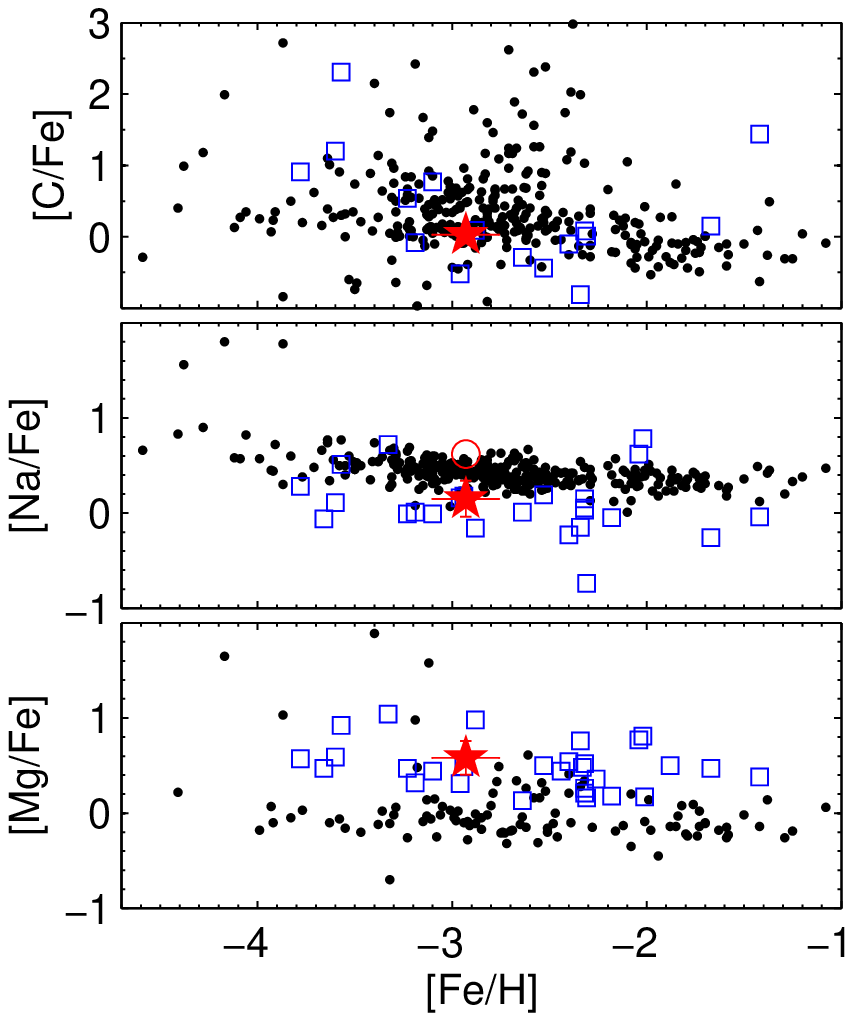}
\includegraphics[angle=0,width=0.5015\hsize]{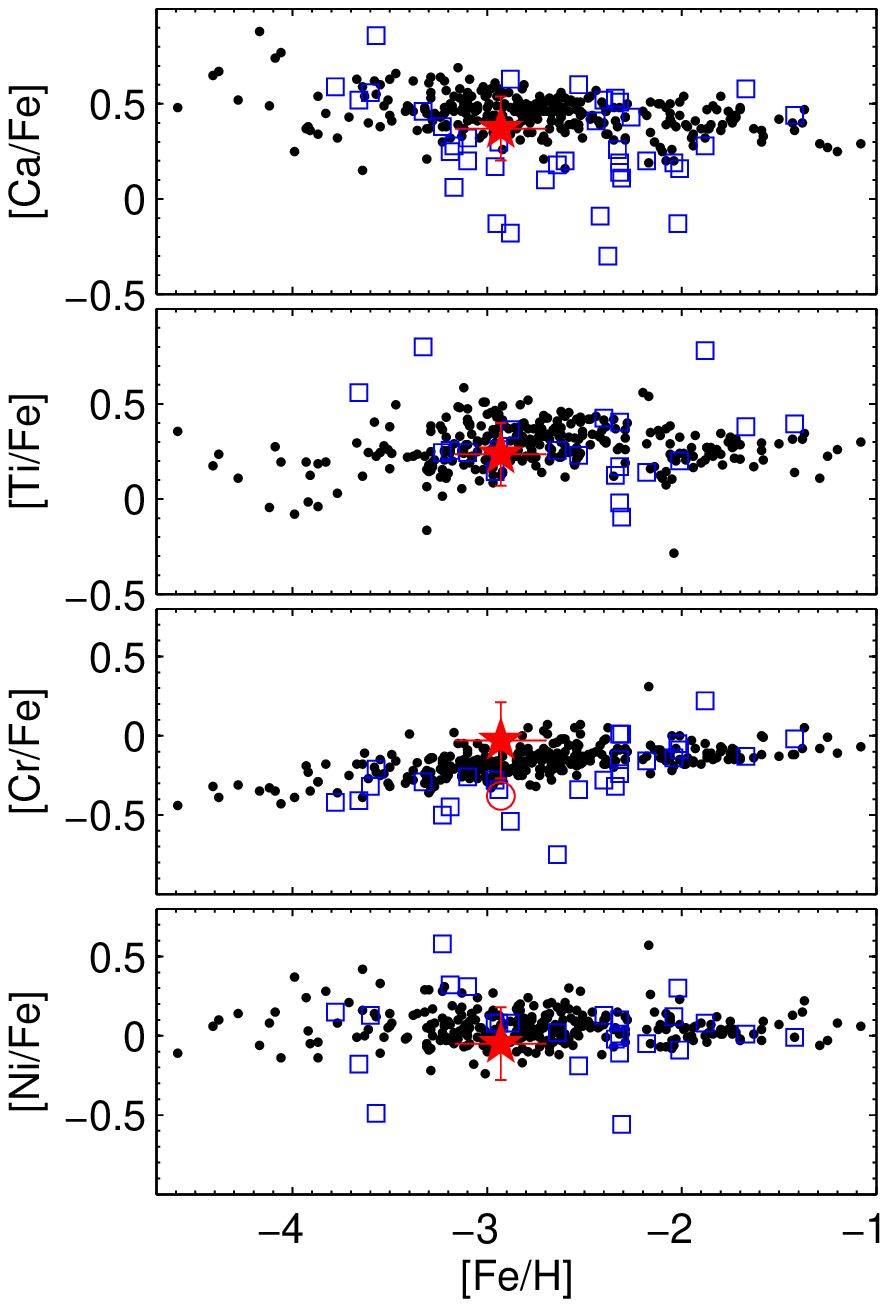}
\end{center}
\caption{Abundance results for C, Na, and Mg (left panels) in Boo~II-15 (red star symbol), metal-poor halo stars (black points; Roederer et al. 2014), and 
seven ultrafaint dSphs (blue squares; see text for references). For the latter, only high-resolution data are shown. 
The right panels show the same samples for the elements Ca, Ti, Cr, and Ni. 
For Na and Cr we show both the NLTE (star) and LTE (open circle) values.
Mind the different abundance-axes ranges between the  panels.}
\end{figure*}
\subsection{Alpha-elements: Mg, Ca, Ti}
None of the commonly used O-features (at 6300 \AA~or the triplet near 7770 \AA) was detectable in our spectrum 
and we attempted to determine upper limits using the Cayrel (1988) formula; these were also consistent with 
the limits obtained by using the formalism by Bohlin et al. (1983). 
However, the upper (1,2,3)$\sigma$ limits of [O/Fe]$<$(1.76, 2.09, 2.30) dex have little meaning in terms of the true chemical enrichment of Boo~II-15 on the lower RGB
and we exclude this element from further discussion. 

Our results for Mg, Ca, and Ti are shown in Fig.~2. 
Thus, the [Mg/Fe] ratio of 0.58 dex is elevated, yet it falls within the broad range covered by the halo abundances. 
Large Mg/Ca ratios have been detected in peculiar stars of some UF dSphs (Koch et al. 2008; Feltzing et al. 2009) and 
also the more luminous dwarf galaxies (Fulbright et al. 2004; Venn et al. 2012). This has been interpreted
by the imprints of the first, massive stars ($\sim$25--50M$_{\odot}$; Koch et al. 2012a) which were able to almost single-handedly carry 
the chemical enrichment of these low-mass systems. 

In contrast, the [Ca/Fe] ratio in Boo~II is not unusually low\footnote{Note, e.g., a population of metal-poor, 
Ca-deficient halo stars,  recently discovered by Cohen et al. (2013).} 
 and is fully compatible with the $\alpha$-enhanced plateau value of the Galactic halo. 

The resulting [Mg/Ca] of 0.32 dex is not outstanding (see also Koch et al. 2009b; Roederer \& Kirby 2014) and does not yield any evidence
for the importance of such massive stars in the evolution of the Boo~II dSph. 

Similarly, the moderately enhanced Ti-abundance in Boo~II-15 falls square on the distribution of halo field stars, while the remainder of 
faint dSphs stars show a broad scatter below [Fe/H]$<-2$ dex. 
Overall, the simple average of Mg, Ca, and Ti suggests an enhancement in this star of [$\alpha$/Fe]$=0.40\pm0.11$ dex.	
While the first measurements of abundances in dSph stars supported Solar [$\alpha$/Fe] ratios, in line with the low star forming efficiencies of these systems 
(e.g., Shetrone et al. 2003), the recent additions of the most metal-poor stars in the (ultra-) faint dSphs   show a broad scatter in this ratio and a strong overlap 
with the metal-poor halo stars (e.g., Koch 2009; Tolstoy et al. 2009). Our finding of a regular $\alpha$-enhancement thus supports the notion that also Boo~II was governed by the same nucleosynthesis 
through SNe~II  as was the halo at early times and, conversely, that the accretion of dSph-like objects \`a la Searle \& Zinn (1978) was  important mostly for the build-up of  
the metal-poor halo at early times. 
\subsection{Light elements: Na}
The only lines strong enough for EW measurements in Boo~II-15 are the resonance NaD lines. 
The Na abundances have been corrected for departures from LTE using data from Lind et al. (2011)\footnote{Taken from the authors' web-based 
database, {\tt www.inspect-stars.net}.} tailored to the stellar parameters of the star.  
This yields NLTE corrections of $-0.49$ and $-0.44$ dex for the two Na  lines.
The  [Na/Fe] ratio of 0.15 dex is thus grazing at the lower edge of the 
 majority of metal-poor halo stars {(also corrected for NLTE)}, while stars in the (UF) dSph stars 
 tend to have slightly systematically lower Na-abundances, in line with our finding in Boo~II-15.  
\subsection{Iron peak: Cr, Ni}
While the [Ni/Fe] ratio is essentially Solar in halo stars over a broad metallicity range (at a mean [Ni/Fe] of 0.04 dex with a 1$\sigma$ scatter of 0.10 dex), 
the values in the UF dSphs show a much broader scatter (mean and 1$\sigma$ are 0.08 and 0.24 dex, respectively). 
Our measurement in Boo~II-15, however, places this star square within the halo and the dSph distributions, leaving no room for peculiarities in the Fe-peak production 
in the environment of this star, {although we caution that the [Ni/Fe] ratio has been derived from a single line.} 

It has been suggested that the subsolar [Cr/Fe] ratios in metal-poor halo stars are a consequence of strong NLTE corrections (e.g., Bergemann \& Cescutti 2010)
or insufficiencies in the SN yields (e.g., McWilliam et al. 1995; Fran\c{c}ois et al. 2004).
Irrespective of either origin, our finding of a low [Cr/Fe] of $-0.38$ dex is fully in line with the low values found in the halo stars. 
Although the calculations by Bergemann \& Cescutti (2010) were made 
for a different range of stellar parameters (dwarfs to subgiants) for a comprehensive  set of Cr\,{\sc i} and {\sc ii} lines, 
in the temperature, gravity, and metallicity regime of Boo~II-15 we expect upward corrections by $\sim$0.35 dex, resulting in [Cr/Fe]$_{\rm NLTE} = -0.03$ dex. 
Bergemann \& Cescutti (2010) conclude  that the  trend of the NLTE-corrected [Cr/Fe] ratios with [Fe/H] is well reproduced by their standard models of Galactic chemical evolution 
without the need for proposing any unusual conditions in the interstellar medium or for altering the nucleosynthetic yields for chromium.  
The fact  that the Cr abundance in Boo~II follows the same, subsolar halo trend 
then also points to a fairly standard enrichment for our target star. 
\subsection{Neutron-capture elements: Ba}
The relatively low S/N ratio of our spectrum prevented us from determining precise abundances from the generally weak lines of most of the heavy elements, be it via their EWs or through spectral synthesis. 
Thus we did not obtain meaningful results for the lighter elements O (though see Sect.~3.6), Al, Si, Sc, V, Mn, Co, Cu, 
nor could we detect any heavy elements such as Zr, Y, or La.
While it is formally possible to compute nominal upper limits for each transition based on the noise characteristics of the spectrum (e.g., Roederer \& Kirby 2014), 
the limits we would thus be able to place on these transitions would have little meaning\footnote{For instance, we can state a limit from the Eu\,{\sc ii} 6645~\AA~line of [Eu/Fe]$<1.90$ dex. 
If Boo~II-15 would indeed have such realistic limits, it would qualify as a strongly $r$-process enhanced very metal-poor star (Sneden et al. 2008), while we do not see any other evidence for 
such  overabundances of the neutron-capture elements in the spectrum.}.

Nevertheless, we chose to pay more attention to Ba, since it is a well-studied representative of the
neutron-capture processes and the four lines accessible in the optical wavelength range (at 4554, 5853, 6141, and 6496~\AA) are usually relatively strong. 
However, in the spectrum of Boo~II-15 we could not detect any of those lines above noise. 
We followed Koch et al. (2013) in determining upper limits  using Cayrel's (1988) formula, accounting for the effects of hyperfine splitting, continuum errors, and 
possible blends.  
If we adopt the upper limit from that line that yields the lowest value  as  the strongest constraint on the Ba-abundance (see also Roederer \& Kirby 2014), 
we can state an upper limit of [Ba/Fe] $< -0.62$ dex (3$\sigma$) from the 4554 \AA~line. 
Fig.~3 illustrates the spectra in the regions around two representative Ba-lines together with synthetic spectra that have been computed with Ba-abundances according to 
the estimated, respective 1, 2, and 3$\sigma$ limits. These syntheses show that Boo~II-15 is indeed characterised by a low abundance, {relative to the Solar value},  of this  neutron-capture element. 
\begin{figure}[!ht]
\begin{center}
\includegraphics[angle=0,width=1\hsize]{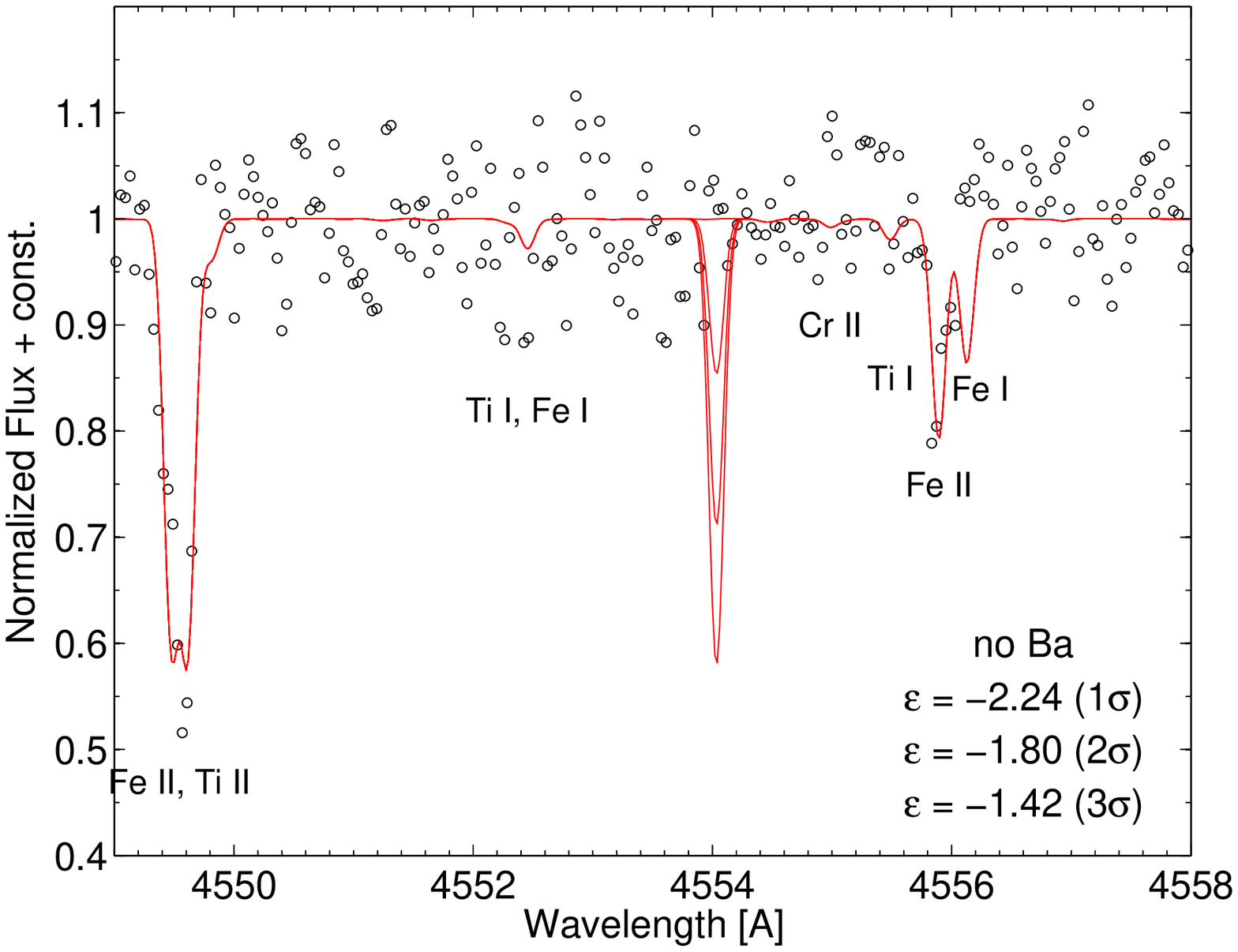}
\includegraphics[angle=0,width=1\hsize]{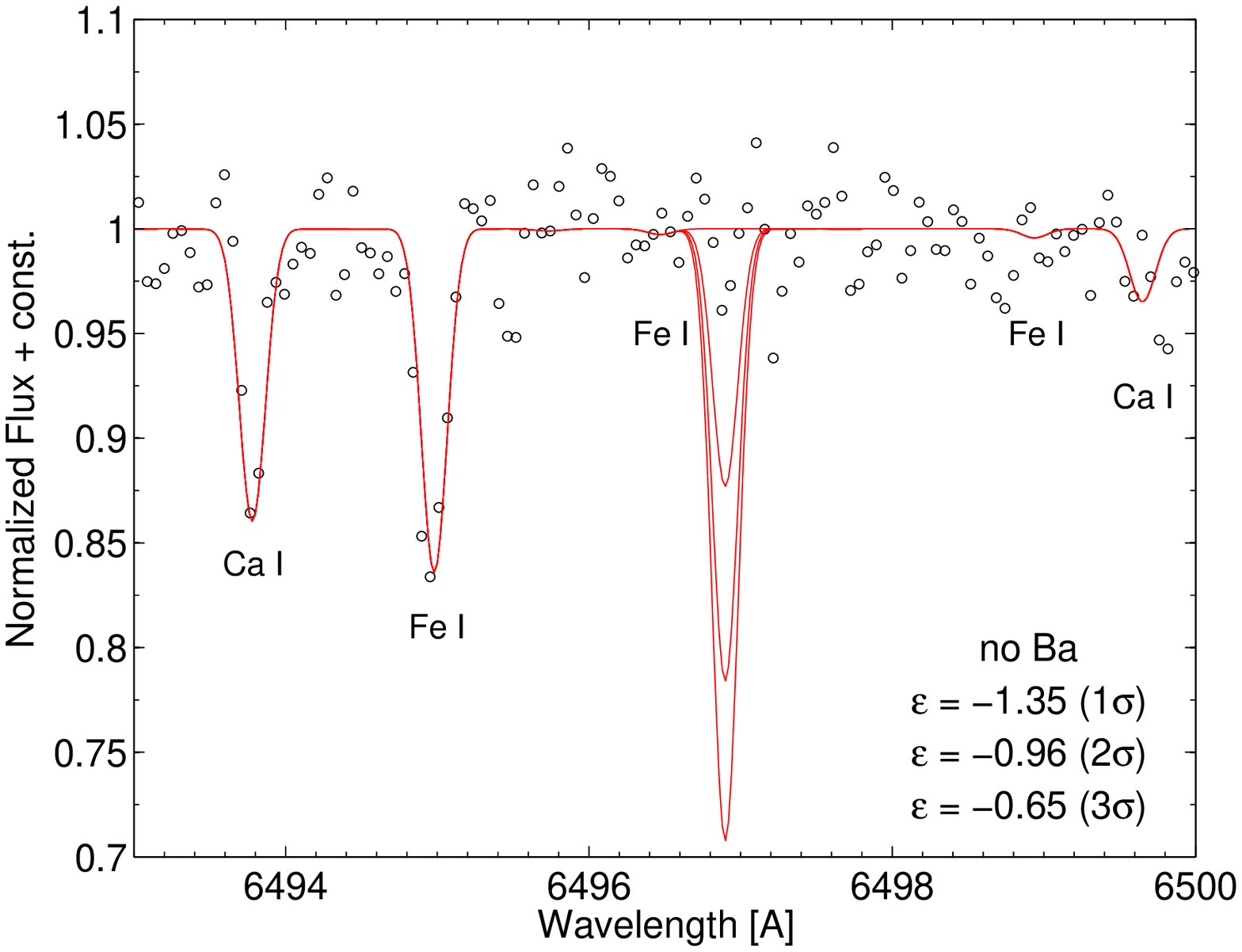}
\end{center}
\caption{Spectra of Boo~II-15 (black circles)  around the Ba\,{\sc ii} lines at 4554.029 \AA~(top panel) and 6496.908 \AA~(bottom). 
Red lines show synthetic spectra that have been computed without any Ba and for the respective 1,2, and 3$\sigma$-limits according to the Cayrel (1988) formula. Other absorption lines 
are also labeled, even if most were too weak for detection as well. The S/N ratios in these regions are $\sim$20 (top) and 40 (bottom) per pixel, respectively.}
\end{figure}

 In Fig.~4 we put this low value in the context of other metal-poor systems.
\begin{figure}[!ht]
\begin{center}
\includegraphics[angle=0,width=1\hsize]{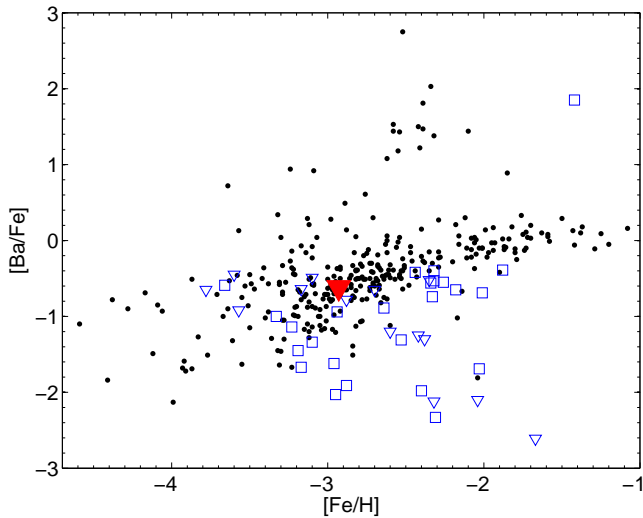}
\end{center}
\caption{Same as Fig.~2 for the abundance ratios of Ba. Upper limits are indicated as triangles.}
\end{figure}
At [Ba/Fe]$<-0.62$ dex, Ba is significantly subsolar in Boo~II-15. Taken at face value, this low abundance ratio is compatible with the metal-poor halo stars and 
the generally strong $n$-capture element deficiencies found in other UF dSphs. Amongst the extremes is Hercules, for which low upper limits on Ba were found over the 
entire metallicity range of more than one dex (Koch et al. 2013). The spectral quality of the present work and the small number statistics of only one target does not allow us to investigate 
any such global depletions in greater detail.

Unfortunately, both standard Sr lines (at 4077 and 4215 \AA) that are commonly used in metal-poor stars' analyses fall outside of our spectral range. 
In combination with Ba, through the Sr/Ba ratio, Sr is a prime  tracer of the earliest enrichment phases of a stellar system, 
in particular for those  stars that are significantly devoid of any of those elements, or in terms of the Galactic chemical evolution of Sr as seen in  EMP halo field stars
(Koch et al. 2013; Roederer 2013; Hansen et al. 2013). 
\section{Reanalysis of the  metallicities from the Calcium triplet}
Traditionally, the analysis of the CaT in red giants was based on {\em linear} calibrations of the line strengths (i.e., a weighted sum of the triplet's individual EWs) 
to stellar metallicity on a reference scale of known iron abundances that uses, amongst others, Galactic GCs (e.g., Rutledge et al. 1997) and/or 
young open clusters (Cole et al. 2004). 
K09 found a mean metallicity of $-1.8$ dex for five stars in Boo~II with a full range of $-2$ to $-1.6$ dex, which is surprisingly high given the galaxy's very low mass.
As elaborated in Sect.~3.4, this may be an overestimate if the Boo~II stars are intrinsically much more metal poor, while our measurements in K09 employed the GC-based calibration of Rutledge et al. (1997), which 
was thus restricted to the more metal-rich regime.
Likewise, due to the very small wavelength coverage of the original (GMOS) spectra and the low S/N ratios for most except the brightest targets of K09, no
other metal lines could be used in the CaT regime for an independent check of the stellar metallicities (as, e.g., in  Hendricks et al. 2014a). 

Recently, Starkenburg et al. (2010) computed a large grid of synthetic CaT spectra and demonstrated that
such linear correlations fail for metal-poor stars below $\sim -3$ dex, where the CaT-metallicity significantly overestimates the true iron abundances. 
By comparison of a large sample of 340 CaT-based metallicities with high-resolution iron abundances in the Fornax dSph, 
Hendricks et al. (2014b) then showed that the GC calibrations overestimate metallicities by up to 0.5 dex below [Fe/H]$<-1.8$ dex. 
While the  empirical calibrations of Starkenburg et al. (2010) provided a better agreement, they still resulted in a systematic offset 
of 0.2 dex over a broad Fe-range, resulting in too metal-rich CaT-metallicities.
Hendricks et al. (2014b) obtained the best match
between high- and low-resolution measurements  in that there were no systematic trends of the residuals ([Fe/H]$_{\rm CaT} - $[Fe/H]$_{\rm HR}$) with metallicity
when using the calibrations of Carrera et al. (2013). 
These authors have reliably 
extended their CaT calibration down to [Fe/H]=$-4$ dex by including a large number of EMP halo stars in their calibrator sample. 
Therefore, we chose to apply this most recent, reliable calibration of Carrera et al. (2013) to our old CaT data from K09. 
The resulting velocity-metallicity diagram is shown in Fig.~5.
\begin{figure}[!ht]
\begin{center}
\includegraphics[angle=0,width=\hsize]{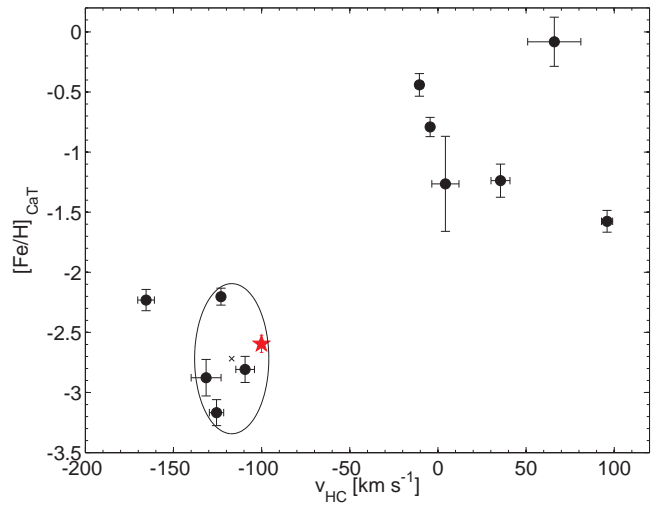}
\end{center}
\caption{Radial velocities and CaT-metallicities derived from the spectra of K09. 
The metallicities have been re-determined using the more recent, reliable  
calibration of Carrera et al. (2013). Boo~II-15 is highlighted as a red star symbol. The {\em bona fide} members of the dSph are the five stars at $-117$ km\,s$^{-1}$ (K09), 
{which we indicate by a 2$\sigma$-dispersion ellipse, with the mean velocity and metallicity plotted by a cross}.}
\end{figure}

The main outcome is now that the sample of five {\em bona fide} member stars in the radial velocity peak of $-117$ km\,s$^{-1}$ becomes more metal poor, with a mean [Fe/H]$_{\rm CaT}$ of $-2.72\pm0.15$ dex
(1$\sigma$ dispersion of 0.3 dex).
Boo~II-15 itself has CaT metallicity of $-2.6\pm0.1$ dex.
The dSph metallicities  cover a full range of $-2.2$ to $-3.2$ dex -- a broad spread that
 is fully in line with other metal-poor, faint dSphs (e.g., Norris et al. 2010a; see also Fig.~1 in Koch et al. 2012b) and that emphasizes that also Boo~II is likely a dark matter dominated system with a potential well that is deep enough to 
 allow for  continuous metal enrichment. 
\section{Discussion}
We carried out the first chemical abundance study of a red giant in the low-luminosity stellar system Boo~II. 
Here, we were able to measure abundances of eight chemical elements and, due to the galaxy's faintness and 
the ensuing low S/N of the spectrum, could only place upper limits on other species.
This star is metal poor, at [Fe/H]=$-2.93$ dex, and shows $\alpha$-enhancements, approximately Solar Fe-peak element abundances, and 
a depletion in the neutron-capture element Ba that are fully consistent with other metal-poor halo field stars and 
dSphs at the same metallicity.

Roederer \& Kirby (2014) analysed a single red giant in Segue~2, the least massive galaxy known to date. 
At essentially the same [Fe/H] as the star of the present study, the two stars in Boo~II and Segue~2 share 
very similar abundance patterns, in particular, very low abundances of the neutron-capture elements. 
{The overall depletions in Ba in several ultrafaint dSphs 
and the subsolar upper limit on [Ba/Fe] we find in Boo~II may pose}
evidence that the
nucleosynthetic channels producing Ba were less efficient in these systems.  Roederer \& Kirby (2014) discuss similarities of the low-mass dSphs with EMP halo stars in that 
some of them have detections of heavy elements, concluding that $n$-capture nucleosynthesis must have taken place already in the first generation of massive, zero-metallicity progenitors.
Based on their chemical abundance analysis, Roederer \& Kirby estimated the total metal content of Segue~2 to be on the order of 0.1 M$_{\odot}$ and argue that such a low
output is  compatible with an enrichment from only one massive  SNe (see also Koch et al. 2008; Simon et al. 2010). 
{Our present abundance limits do not allow for a quantitative  
 assessment of Boo~II's enrichment history; however,  based on its low stellar mass and the subsolar $n$-capture element abundance 
 we cannot exclude the possibility that 
 individual SNe events may have played an important role in its evolution, as seems to be common amongst the UF dSphs.}

Based on Boo~II's overlap in projection and in radial velocity with the Sagittarius Stream, and given its overall morphology, its CMD, and kinematics, K09 
discussed the possibility that this overdensity may not be an old and metal-poor dSph, but could also be a dissolved star cluster or disrupted dSph formerly associated with Sgr. 
However, with the present chemical abundance analysis we have now efficiently shown that this is an unlikely scenario and that all evidence hints at Boo~II being indeed 
a purely old and metal-poor system.

Neither the Sgr field star and stream population extend as metal poor as the metallicities present in Boo~II (Vivas et al. 2005; Monaco et al. 2007), 
nor are any of the GCs associated with Sgr that metal poor (Law \& Majewski 2010). Moreover, the confirmation of an abundance spread in this system provides 
clear evidence in favor of a dark-matter dominated satellite with continuous enrichment. 
The fact that these systems have abundance spreads in excess of one dex also suggests that 
 the few stars that can be observed probably only permit a mere glimpse into the true, broad range covered by this dSph.

Assuming the low absolute magnitude for Boo~II  of M$_V=-2.7$ mag,  or $1.0\pm0.8$$\times$$10^3$ L$_{\odot}$ (Walsh et al. 2007, 2008; Martin et al. 2008; K09), 
the linear luminosity-metallicity relation of Kirby et al. (2011), which  holds over about  four orders of magnitude,
 predicts a mean metallicity for this UF dSph of $-2.65\pm0.15$ dex. This is fully consistent with the metal-poor mean of Boo~II we find from a re-calibration of the K09 CaT spectra using 
 the new calibration of Carrera et al. (2013). %
Our finding thus shows that Boo~II's metallicity distribution is indeed consistent with its present-day luminosity, and thus mass, so that tidal stripping was probably not an important 
factor in its early evolution (cf. Roederer \& Kirby 2014). 
Finally we note that the new mean metallicity of Bo\"otes~II from the reliable calibration of the CaT is significantly lower than the early estimate from K09, based on traditional
 GC-based relations, which demonstrates that the choice of calibrations in the metal-poor regime is of prime importance when assessing the characteristics of these faint,  stellar systems. 
\begin{acknowledgments}
We thank the anonymous referee for a a prompt and helpful report. 
AK acknowledges the Deutsche Forschungsgemeinschaft for funding from  Emmy-Noether grant  Ko 4161/1. 
RMR acknowledges support from grants AST-1212095 and 1413755 from the National Science Foundation. 
This research made use of atomic  data from the INSPECT database, version 1.0 (www.inspect-stars.net). 
The data presented herein were obtained at the W.M. Keck Observatory, which is operated as a scientific partnership 
among the California Institute of Technology, the University of California and the National Aeronautics and Space Administration. 
The Observatory was made possible by the generous financial support of the W.M. Keck Foundation.
\end{acknowledgments}
\end{document}